\PassOptionsToPackage{unicode}{hyperref}
\PassOptionsToPackage{hyphens}{url}
\PassOptionsToPackage{dvipsnames,svgnames,x11names}{xcolor}
\documentclass[
]{article}
\usepackage{amsmath,amssymb}
\usepackage{lmodern}
\usepackage{iftex}
\ifPDFTeX
  \usepackage[T1]{fontenc}
  \usepackage[utf8]{inputenc}
  \usepackage{textcomp} 
\else 
  \usepackage{unicode-math}
  \defaultfontfeatures{Scale=MatchLowercase}
  \defaultfontfeatures[\rmfamily]{Ligatures=TeX,Scale=1}
\fi
\IfFileExists{upquote.sty}{\usepackage{upquote}}{}
\IfFileExists{microtype.sty}{
  \usepackage[]{microtype}
  \UseMicrotypeSet[protrusion]{basicmath} 
}{}
\makeatletter
\@ifundefined{KOMAClassName}{
  \IfFileExists{parskip.sty}{%
    \usepackage{parskip}
  }{
    \setlength{\parindent}{0pt}
    \setlength{\parskip}{6pt plus 2pt minus 1pt}}
}{
  \KOMAoptions{parskip=half}}
\makeatother
\usepackage{xcolor}
\usepackage{longtable,booktabs,array}
\usepackage{calc} 
\usepackage{etoolbox}
\makeatletter
\patchcmd\longtable{\par}{\if@noskipsec\mbox{}\fi\par}{}{}
\makeatother
\IfFileExists{footnotehyper.sty}{\usepackage{footnotehyper}}{\usepackage{footnote}}
\makesavenoteenv{longtable}
\usepackage{graphicx}
\makeatletter
\def\maxwidth{\ifdim\Gin@nat@width>\linewidth\linewidth\else\Gin@nat@width\fi}
\def\maxheight{\ifdim\Gin@nat@height>\textheight\textheight\else\Gin@nat@height\fi}
\makeatother
\setkeys{Gin}{width=\maxwidth,height=\maxheight,keepaspectratio}
\makeatletter
\def\fps@figure{htbp}
\makeatother
\newlength{\cslhangindent}
\newlength{\csllabelwidth}
\newlength{\cslentryspacingunit} 
\newenvironment{CSLReferences}[2] 
 {
  \setlength{\parindent}{0pt}
  \ifodd #1
  \let\oldpar\par
  \def\par{\hangindent=\cslhangindent\oldpar}
  \fi
  \setlength{\parskip}{#2\cslentryspacingunit}
 }%
 {}
\usepackage{calc}

\ifLuaTeX
\usepackage[bidi=basic]{babel}
\else
\usepackage[bidi=default]{babel}
\fi
\babelprovide[main,import]{american}

\def\languageshorthands#1{}
\ifLuaTeX
  \usepackage{selnolig}  
\fi
\IfFileExists{bookmark.sty}{\usepackage{bookmark}}{\usepackage{hyperref}}
\IfFileExists{xurl.sty}{\usepackage{xurl}}{} 
\hypersetup{
  pdftitle={EXP: a Python/C++ package for basis function expansion
methods in galactic dynamics},
  pdfauthor={Michael S. Petersen, Martin D. Weinberg},
  pdflang={en-US},
  colorlinks=true,
  linkcolor={Maroon},
  filecolor={Maroon},
  citecolor={Blue},
  urlcolor={Blue},
  pdfcreator={LaTeX via pandoc}}

\title{EXP: a Python/C++ package for basis function expansion methods in
galactic dynamics}

\usepackage[affil-it]{authblk}
\usepackage{orcidlink}
\author[1%
  ]{Michael S. Petersen%
    \,\orcidlink{0000-0003-1517-3935}\,%
    }
\author[2%
  ]{Martin D. Weinberg%
    \,\orcidlink{0000-0003-2660-2889}\,%
    }

\affil[1]{University of Edinburgh, UK}
\affil[2]{University of Massachusetts Amherst, USA}
\date{09 May 2025}

\begin{document}
\maketitle

\hypertarget{summary}{%
\section{Summary}\label{summary}}

Galaxies are ensembles of dark and baryonic matter confined by their
mutual gravitational attraction. The dynamics of galactic evolution have
been studied with a combination of numerical simulations and analytical
methods (\protect\hyperlink{ref-Binney:2008}{Binney \& Tremaine, 2008})
for decades. Recently, data describing the positions and motions of
billions of stars from the \emph{Gaia satellite}
(\protect\hyperlink{ref-gaia}{Gaia Collaboration, 2016},
\protect\hyperlink{ref-gaia_DR2_disk}{2018}) depict a Milky Way (our
home galaxy) much farther from equilibrium than we imagined and beyond
the range of many analytic models. Simulations that allow collections of
bodies to evolve under their mutual gravity are capable of reproducing
such complexities but robust links to fundamental theoretical
explanations are still missing.

Basis Function Expansions (BFEs) represent fields as a linear
combination of orthogonal functions. BFEs are particularly well-suited
for studies of perturbations from equilibrium, such as the evolution of
a galaxy. For any galaxy simulation, a biorthogonal BFE can fully
represent the density, potential and forces by time series of
coefficients. The coefficients have physical meaning: they represent the
gravitational potential energy in a given function. The variation of the
function coefficients in time encodes the dynamical evolution. The
representation of simulation data by BFEs results in huge compression of
the information in the dynamical fields. For example, 1.5 TB of phase
space data enumerating the positions and velocities of millions of
particles becomes 200 MB of coefficient data!

\begin{figure}
\centering
\includegraphics{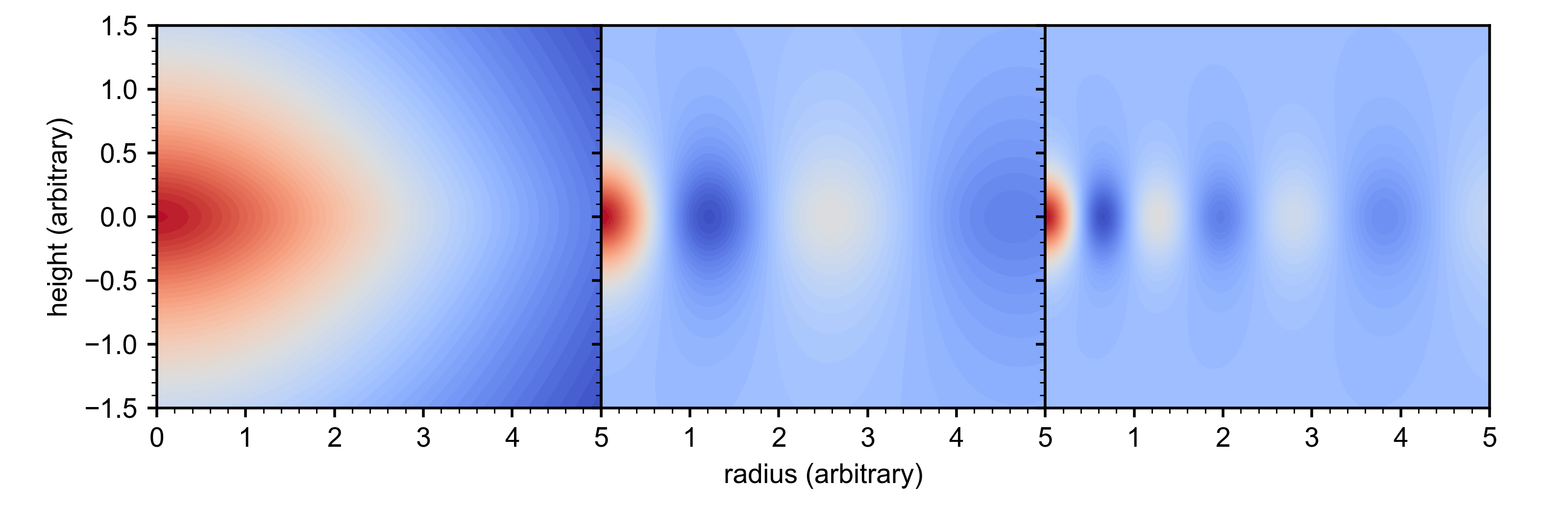}
\caption{Example cylinder basis functions, where the color encodes the
amplitude of the function, for an exponential disk with a scale length
of 3 and a scale height of 0.3 in arbitrary units. We select three
functions at low, medium, and higher order (corresponding to the number
of nodes). The color scale has been normalised such that the largest
amplitude is unity in each panel. \label{fig:examplecylinder}}
\end{figure}

For optimal representation, the lowest-order basis function should be
similar to the mean or equilibrium profile of the galaxy. This allows
for use of the fewest number of terms in the expansion. For example, the
often-used basis set from Hernquist \& Ostriker
(\protect\hyperlink{ref-Hernquist:92}{1992}) matches the Hernquist
(\protect\hyperlink{ref-Hernquist:90}{1990}) dark matter halo profile,
yet this basis set is inappropriate for representing the
cosmologically-motivated Navarro et al.
(\protect\hyperlink{ref-NFW}{1997}) profile. The \texttt{EXP} software
package implements the adaptive empirical orthogonal function (EOF; see
\autoref{fig:examplecylinder}) basis strategy originally described in
Weinberg (\protect\hyperlink{ref-Weinberg:99}{1999}) that matches any
physical system close to an equilibrium model. The package includes both
a high performance N-body simulation toolkit with computational effort
scaling linearly with N (\protect\hyperlink{ref-Petersen:22}{Petersen et
al., 2022}), and a user-friendly Python interface called \texttt{pyEXP}
that enables BFE and time-series analysis of any N-body simulation
dataset.

\hypertarget{statement-of-need}{%
\section{Statement of need}\label{statement-of-need}}

The need for methodology that seamlessly connects theoretical
descriptions of dynamics, N-body simulations, and compact descriptions
of observed data gave rise to \texttt{EXP}. This package provides recent
developments from applied mathematics and numerical computation to
represent complete series of BFEs that describe the variation of
\emph{any} field in space. In the context of galactic dynamics, these
fields may be density, potential, force, velocity fields or any
intrinsic field produced by simulations such as chemistry data. By
combining the coefficient information through time using multichannel
singular spectral analysis (mSSA, \protect\hyperlink{ref-SSA}{Golyandina
et al., 2001}), a non-parametric spectral technique, \texttt{EXP} can
deepen our understanding by discovering the dynamics of galaxy evolution
directly from simulated, and by analogy, observed data.

\texttt{EXP} decomposes a galaxy into multiple bases for a variety of
scales and geometries and is thus able to represent arbitrarily complex
simulations with many components (e.g., disk, bulge, dark matter halo,
satellites). \texttt{EXP} is able to efficiently summarize the degree
and nature of asymmetries through coefficient amplitudes tracked through
time and provide details at multiple scales. The amplitudes themselves
enable ex-post-facto dynamical discovery. \texttt{EXP} is a collection
of object-oriented C++ libraries with an associated modular N-body code
and a suite of stand-alone analysis applications.

\texttt{pyEXP} provides a full Python interface to the \texttt{EXP}
libraries, implemented with \texttt{pybind11}
(\protect\hyperlink{ref-pybind11}{Jakob et al., 2017}), which provides
full interoperability with major astronomical packages including Astropy
(\protect\hyperlink{ref-astropy}{Astropy Collaboration, 2013}) and
\texttt{Gala} (\protect\hyperlink{ref-gala}{Price-Whelan, 2017}).
Example workflows based on previously published work are available and
distributed as accompanying
\href{https://github.com/EXP-code/pyEXP-examples}{examples and
tutorials}. The examples and tutorials flatten the learning curve for
adopting BFE tools to generate and analyze the significance of
coefficients and discover dynamical relationships using time series
analysis such as mSSA. We provide a
\href{https://exp-docs.readthedocs.io}{full online manual} hosted by
ReadTheDocs.

The software package brings published---but difficult to
implement---applied-math technologies into the astronomical mainstream.
\texttt{EXP} and the associated Python interface \texttt{pyEXP}
accomplish this by providing tools integrated with the Python ecosystem,
and in particular are well-suited for interactive Python
(\protect\hyperlink{ref-iPython}{Pérez \& Granger, 2007}) use through,
e.g., Jupyter notebooks (\protect\hyperlink{ref-jupyter}{Kluyver et al.,
2016}). \texttt{EXP} serves as the scaffolding for new imaginative
applications in galactic dynamics, providing a common dynamical language
for simulations and analytic theory.

\hypertarget{features-and-workflow}{%
\section{Features and workflow}\label{features-and-workflow}}

The core \texttt{EXP} library is built around methods to build the best
basis function expansion for an arbitrary data set in galactic dynamics.
The table below lists some of the available basis functions. All
computed bases and resulting coefficient data are stored in HDF5
(\protect\hyperlink{ref-hdf5}{The HDF Group, 2000-2010}) format.

\begin{longtable}[]{@{}
  >{\raggedright\arraybackslash}p{(\columnwidth - 2\tabcolsep) * \real{0.4583}}
  >{\raggedright\arraybackslash}p{(\columnwidth - 2\tabcolsep) * \real{0.5417}}@{}}
\toprule\noalign{}
\begin{minipage}[b]{\linewidth}\raggedright
Name
\end{minipage} & \begin{minipage}[b]{\linewidth}\raggedright
Description
\end{minipage} \\
\midrule\noalign{}
\endhead
\bottomrule\noalign{}
\endlastfoot
sphereSL & Sturm--Liouville basis function solutions to Poisson's
equation for any arbitrary input spherical density \\
bessel & Basis constructed from eigenfunctions of the spherical
Laplacian \\
cylinder & EOF solutions tabulated on the meridional plane for
distributions with cylindrical geometries \\
flatdisk & EOF basis solutions for the three-dimensional gravitational
field of a razor-thin disk \\
cube & Trigonometric basis solution for expansions in a cube with
boundary criteria \\
field & General-purpose EOF solution for scalar profiles \\
velocity & EOF solution for velocity flow coefficients \\
\end{longtable}

\hypertarget{n-body-simulation}{%
\subsection{N-body simulation}\label{n-body-simulation}}

Computing the gravitational potential and forces from a collection of N
particles is typically an expensive endeavour. \texttt{EXP} reduces the
cost by using BFE to compute the potential and forces such that
computational effort scales with the number of particles. Other modern
N-body codes use direct summation (\protect\hyperlink{ref-Wang:15}{Wang
et al., 2015}) or tree-based solutions
(\protect\hyperlink{ref-Gadget4}{Springel et al., 2021}), which have
computational effort that scales as N\(^2\) and N log N, respectively.
The trade off for BFE solutions comes in the form of restricted degrees
of freedom. For many problems in near-equilibrium galactic dynamics this
is not a problem, but rather a feature.

Our design includes a wide choice of run-time summary diagnostics,
phase-space output formats, dynamically loadable user libraries, and
easy extensibility. Stand-alone routines include the EOF and mSSA
methods described above, and the modular software architecture of
\texttt{EXP} enables users to easily build and maintain extensions. The
\texttt{EXP} code base is described in published papers
(\protect\hyperlink{ref-Petersen:22}{Petersen et al., 2022};
\protect\hyperlink{ref-Weinberg:23}{Weinberg, 2023}) and has been used,
enhanced, and rigorously tested for nearly two decades.

The design and implementation of the N-body tools allows for execution
on a wide variety of hardware, from personal laptops to high performance
computing centers, with communication between processes handled by MPI
(\protect\hyperlink{ref-mpi41}{Message Passing Interface Forum, 2023})
and GPU implementations in CUDA (\protect\hyperlink{ref-cuda}{NVIDIA et
al., 2020}). Owing to the linear scaling of computational effort with N
and the novel GPU implementation, the N-body methods in \texttt{EXP}
deliver performance in collisionless N-body simulations previously only
accessible with large dedicated CPU clusters.

The flexible N-body software design allows users to write their own
modules for on-the-fly execution during N-body integration. Modules
enable powerful and intricate dynamical experiments in N-body
simulations, further reducing the gap between numerical simulations and
analytic dynamics. The package ships with several examples, including
imposed external potentials, as well as a basic example that can be
extended by users.

\hypertarget{using-pyexp-to-represent-simulations}{%
\subsection{Using pyEXP to represent
simulations}\label{using-pyexp-to-represent-simulations}}

\texttt{pyEXP} provides an interface to many of the classes in the
\texttt{EXP} C++ library, allowing for both the generation of all bases
listed in the table above as well as coefficients for an input data set.
Each of these tools are Python classes that accept NumPy
(\protect\hyperlink{ref-numpy}{Harris et al., 2020}) arrays for
immediate interoperability with Matplotlib
(\protect\hyperlink{ref-matplotlib}{Hunter, 2007}) and Astropy. We
include a verified set of stand-alone routines that read phase-space
files from many major cosmological tree codes (e.g.,
\protect\hyperlink{ref-Gadget4}{Springel et al., 2021}) and produce
BFE-based analyses. The code suite includes adapters for reading and
writing phase space for many of the widely used cosmology codes, with a
base class for developing new ones. There are multiple ways to use the
versatile and modular tools in \texttt{pyEXP}, and we anticipate
pipelines that we have not yet imagined. The flexibility of the basis
sets available in \texttt{EXP} greatly enhances the number of available
basis sets implemented in Python (see, e.g.
\protect\hyperlink{ref-gala}{Price-Whelan, 2017}).

\hypertarget{using-pyexp-to-analyze-time-series}{%
\subsection{Using pyEXP to analyze time
series}\label{using-pyexp-to-analyze-time-series}}

The \texttt{EXP} library includes multiple time series analysis tools,
documented in the manual. Here, we briefly highlight one technique that
we have already used in published work: mSSA
(\protect\hyperlink{ref-Johnson:23}{Johnson et al., 2023};
\protect\hyperlink{ref-Weinberg:21}{Weinberg \& Petersen, 2021}).
Beginning with coefficient series from the previous tools, mSSA
summarizes signals \emph{in time} that describe dynamically correlated
responses and patterns. Essentially, this is BFE in time and space.
These temporal and spatial patterns allow users to better identify
dynamical mechanisms and enable intercomparisons and filtering for
features in simulation suites, e.g.~computing the fraction galaxies with
grand design structure or hosting bars. Random-matrix techniques for
singular-value decomposition ensure that analyses of large data sets is
possible. All mSSA decompositions are saved in HDF5 format for reuse.

\hypertarget{acknowledgements}{%
\section{Acknowledgements}\label{acknowledgements}}

We acknowledge the ongoing support of the \href{https://exp-code.github.io/}{EXP
collaboration}. We also acknowledge the support of the Center for
Computational Astrophysics (CCA). The CCA is part of the Flatiron
Institute, funded by the Simons Foundation. We thank Robert Blackwell
for invaluable help with HPC best practices.

\hypertarget{references}{%
\section*{References}\label{references}}
\addcontentsline{toc}{section}{References}

\hypertarget{refs}{}
\begin{CSLReferences}{1}{0}
\leavevmode\vadjust pre{\hypertarget{ref-astropy}{}}%
Astropy Collaboration. (2013). {Astropy: A community Python package for
astronomy}. \emph{Astronomy \& Astrophysics}, \emph{558}.
\url{https://doi.org/10.1051/0004-6361/201322068}

\leavevmode\vadjust pre{\hypertarget{ref-Binney:2008}{}}%
Binney, J., \& Tremaine, S. (2008). \emph{{Galactic Dynamics: Second
Edition}}. Princeton University Press.
\url{http://adsabs.harvard.edu/abs/2008gady.book.....B}

\leavevmode\vadjust pre{\hypertarget{ref-gaia}{}}%
Gaia Collaboration. (2016). {The Gaia mission}. \emph{Astronomy \&
Astrophysics}, \emph{595}, A1.
\url{https://doi.org/10.1051/0004-6361/201629272}

\leavevmode\vadjust pre{\hypertarget{ref-gaia_DR2_disk}{}}%
Gaia Collaboration. (2018). {Gaia Data Release 2. Mapping the Milky Way
disc kinematics}. \emph{Astronomy \& Astrophysics}, \emph{616}, A11.
\url{https://doi.org/10.1051/0004-6361/201832865}

\leavevmode\vadjust pre{\hypertarget{ref-SSA}{}}%
Golyandina, N., Nekrutkin, V., \& Zhigljavsky, A. A. (2001).
\emph{Analysis of time series structure: SSA and related techniques}.
CRC press.

\leavevmode\vadjust pre{\hypertarget{ref-numpy}{}}%
Harris, C. R., Millman, K. J., Walt, S. J. van der, Gommers, R.,
Virtanen, P., Cournapeau, D., Wieser, E., Taylor, J., Berg, S., Smith,
N. J., Kern, R., Picus, M., Hoyer, S., Kerkwijk, M. H. van, Brett, M.,
Haldane, A., Río, J. F. del, Wiebe, M., Peterson, P., \ldots{} Oliphant,
T. E. (2020). Array programming with {NumPy}. \emph{Nature},
\emph{585}(7825), 357--362.
\url{https://doi.org/10.1038/s41586-020-2649-2}

\leavevmode\vadjust pre{\hypertarget{ref-Hernquist:90}{}}%
Hernquist, L. (1990). An analytical model for spherical galaxies and
bulges. \emph{The Astrophysical Journal}, \emph{356}, 359.
\url{https://doi.org/10.1086/168845}

\leavevmode\vadjust pre{\hypertarget{ref-Hernquist:92}{}}%
Hernquist, L., \& Ostriker, J. P. (1992). A self-consistent field method
for galactic dynamics. \emph{The Astrophysical Journal}, \emph{386},
375. \url{https://doi.org/10.1086/171025}

\leavevmode\vadjust pre{\hypertarget{ref-matplotlib}{}}%
Hunter, J. D. (2007). Matplotlib: A 2D graphics environment.
\emph{Computing in Science \& Engineering}, \emph{9}(3), 90--95.
\url{https://doi.org/10.1109/MCSE.2007.55}

\leavevmode\vadjust pre{\hypertarget{ref-pybind11}{}}%
Jakob, W., Rhinelander, J., \& Moldovan, D. (2017). \emph{{pybind11 --
Seamless operability between C++11 and Python}}.

\leavevmode\vadjust pre{\hypertarget{ref-Johnson:23}{}}%
Johnson, A. C., Petersen, M. S., Johnston, K. V., \& Weinberg, M. D.
(2023). {Dynamical data mining captures disc-halo couplings that
structure galaxies}. \emph{Monthly Notices of the Royal Astronomical
Society}, \emph{521}(2), 1757--1774.
\url{https://doi.org/10.1093/mnras/stad485}

\leavevmode\vadjust pre{\hypertarget{ref-jupyter}{}}%
Kluyver, T., Ragan-Kelley, B., Pérez, F., Granger, B., Bussonnier, M.,
Frederic, J., Kelley, K., Hamrick, J., Grout, J., Corlay, S., Ivanov,
P., Avila, D., Abdalla, S., Willing, C., \& Jupyter Development Team.
(2016). {Jupyter Notebooks{\textemdash}a publishing format for
reproducible computational workflows}. In \emph{IOS press} (pp. 87--90).
\url{https://doi.org/10.3233/978-1-61499-649-1-87}

\leavevmode\vadjust pre{\hypertarget{ref-mpi41}{}}%
Message Passing Interface Forum. (2023). \emph{{MPI}: A message-passing
interface standard version 4.1}.
\url{https://www.mpi-forum.org/docs/mpi-4.1/mpi41-report.pdf}

\leavevmode\vadjust pre{\hypertarget{ref-NFW}{}}%
Navarro, J. F., Frenk, C. S., \& White, S. D. M. (1997). A universal
density profile from hierarchical clustering. \emph{The Astrophysical
Journal}, \emph{490}(2), 493--508. \url{https://doi.org/10.1086/304888}

\leavevmode\vadjust pre{\hypertarget{ref-cuda}{}}%
NVIDIA, Vingelmann, P., \& Fitzek, F. H. P. (2020). \emph{CUDA, release:
10.2.89}. \url{https://developer.nvidia.com/cuda-toolkit}

\leavevmode\vadjust pre{\hypertarget{ref-iPython}{}}%
Pérez, F., \& Granger, B. E. (2007). {IP}ython: A system for interactive
scientific computing. \emph{Computing in Science and Engineering},
\emph{9}(3), 21--29. \url{https://doi.org/10.1109/MCSE.2007.53}

\leavevmode\vadjust pre{\hypertarget{ref-Petersen:22}{}}%
Petersen, M. S., Weinberg, M. D., \& Katz, N. (2022). {EXP: N-body
integration using basis function expansions}. \emph{Monthly Notices of
the Royal Astronomical Society}, \emph{510}(4), 6201--6217.
\url{https://doi.org/10.1093/mnras/stab3639}

\leavevmode\vadjust pre{\hypertarget{ref-gala}{}}%
Price-Whelan, A. M. (2017). {Gala: A Python package for galactic
dynamics}. \emph{The Journal of Open Source Software}, \emph{2}(18).
\url{https://doi.org/10.21105/joss.00388}

\leavevmode\vadjust pre{\hypertarget{ref-Gadget4}{}}%
Springel, V., Pakmor, R., Zier, O., \& Reinecke, M. (2021). {Simulating
cosmic structure formation with the GADGET-4 code}. \emph{Monthly
Notices of the Royal Astronomical Society}, \emph{506}(2), 2871--2949.
\url{https://doi.org/10.1093/mnras/stab1855}

\leavevmode\vadjust pre{\hypertarget{ref-hdf5}{}}%
The HDF Group. (2000-2010). \emph{{Hierarchical Data Format, version
5}}. http://www.hdfgroup.org/HDF5.

\leavevmode\vadjust pre{\hypertarget{ref-Wang:15}{}}%
Wang, L., Spurzem, R., Aarseth, S., Nitadori, K., Berczik, P.,
Kouwenhoven, M. B. N., \& Naab, T. (2015). {NBODY6++GPU: ready for the
gravitational million-body problem}. \emph{Monthly Notices of the Royal
Astronomical Society}, \emph{450}(4), 4070--4080.
\url{https://doi.org/10.1093/mnras/stv817}

\leavevmode\vadjust pre{\hypertarget{ref-Weinberg:99}{}}%
Weinberg, M. D. (1999). An adaptive algorithm for {N}-body field
expansions. \emph{The Astronomical Journal}, \emph{117}(1), 629--637.
\url{https://doi.org/10.1086/300669}

\leavevmode\vadjust pre{\hypertarget{ref-Weinberg:23}{}}%
Weinberg, M. D. (2023). {New dipole instabilities in spherical stellar
systems}. \emph{Monthly Notices of the Royal Astronomical Society},
\emph{525}(4), 4962--4975. \url{https://doi.org/10.1093/mnras/stad2591}

\leavevmode\vadjust pre{\hypertarget{ref-Weinberg:21}{}}%
Weinberg, M. D., \& Petersen, M. S. (2021). {Using multichannel singular
spectrum analysis to study galaxy dynamics}. \emph{Monthly Notices of
the Royal Astronomical Society}, \emph{501}(4), 5408--5423.
\url{https://doi.org/10.1093/mnras/staa3997}

\end{CSLReferences}

\end{document}